\documentclass[10pt, conference]{IEEEtran}

\makeatletter
\def\ps@headings{%
\def\@oddhead{\mbox{}\scriptsize\rightmark \hfil \thepage}%
\def\@evenhead{\scriptsize\thepage \hfil \leftmark\mbox{}}%
\def\@oddfoot{}
\def\@evenfoot{}}
\def\blfootnote{\xdef\@thefnmark{}\@footnotetext}
\makeatother
\usepackage{graphicx} 
\usepackage{epsfig} 
\usepackage{color}
\usepackage{mathptmx} 
\usepackage{times} 
\usepackage{latexsym}
\usepackage{amsmath} 
\usepackage{amssymb}  
\usepackage{subfigure}
\usepackage{algorithmic}
\usepackage{setspace}
\usepackage[bottom]{footmisc}
\newtheorem{theorem}{Theorem}

\newtheorem{definition}{Definition}
\newtheorem{corollary}{Corollary}

\title{\huge \bf
Optimal strategies for computing symmetric Boolean functions in collocated networks
}

\author{\IEEEauthorblockN{Hemant Kowshik}
\IEEEauthorblockA{CSL and Department of ECE\\
University of Illinois Urbana-Champaign\\
Email: kowshik2@illinois.edu}
\and
\IEEEauthorblockN{P. R. Kumar}
\IEEEauthorblockA{CSL and Department of ECE\\
University of Illinois Urbana-Champaign\\
Email: prkumar@illinois.edu}
}
\begin{document}
\maketitle\blfootnote{This material is based upon work partially supported by AFOSR under
Contract FA9550-09-0121, NSF under Contract Nos. CNS-05-19535, CNS-07-21992, ECCS-0701604, and CNS-0626584, and USARO under Contract Nos. W911NF-08-1-0238 and W-911-NF-0710287.}
\thispagestyle{empty}
\pagestyle{empty}
\begin{abstract}
We address the problem of finding optimal strategies for computing Boolean symmetric functions. We consider a collocated network, where each node's transmissions can be heard by every other node. Each node has a Boolean measurement and we wish to compute a given Boolean function of these measurements with zero error. We allow for block computation to enhance data fusion efficiency, and determine the minimum worst-case total bits to be communicated to perform the desired computation. We restrict attention to the class of symmetric Boolean functions, which only depend on the number of $1$s among the $n$ measurements.  

We define three classes of functions, namely threshold functions, delta functions and interval functions. We provide exactly optimal strategies for the first two classes, and an order-optimal strategy with optimal preconstant for interval functions. Using these results, we can characterize the complexity of computing \textit{percentile} type functions, which is of great interest. In our analysis, we use lower bounds from communication complexity theory, and provide an achievable scheme using information theoretic tools. 
\end{abstract}
\section{INTRODUCTION}
Wireless sensor networks are composed of nodes with sensing, wireless communication and computation capabilities. These networks are designed for applications like fault monitoring, data harvesting and environmental monitoring. In these applications, one is interested  only in computing some relevant \textit{function} of the measurements. For example, one might want to compute the mean temperature for environmental monitoring, or the maximum temperature in fire alarm systems. This suggests moving away from a data forwarding paradigm, and focusing on efficient in-network computation and communication strategies for the function of interest. 


The problem of computing functions of distributed data in sensor networks presents several challenges. On the one hand, the wireless medium being a broadcast medium, nodes have to deal with interference from other transmissions. On the other hand, nodes can exploit these overheard transmissions, and the structure of the function to be computed, to achieve a more efficient description of their own data. This is a sigificant departure from the traditional decode and forward paradigm. 

We consider a collocated network where each node's transmissions can be heard by every other node. This could correspond to a collocated subnet in a sensor network. Each node has a Boolean variable and we focus on the specific problem of symmetric Boolean function computation. We adopt a deterministic formulation of the problem of function computation, allowing zero error. We consider the problem of worst-case function computation, without imposing a probability distribution on the node measurements. Further, instead of restricting a strategy to compute just one instance of the problem, we allow for a block of $N$ independent instances for which the function is to be computed. Thus nodes can accumulate a block of $N$ measurements, and realize greater efficiency by using block codes. 

We assume a packet capture model as in \cite{GiridharKumar}, where collisions do not convey information. Thus, the problem of medium access is resolved by allowing at most one node to transmit successfully at any time. The set of admissible strategies includes all interactive strategies, where a node may exchange several messages with other nodes. 
It is of particular interest to study the benefit of interactive strategies versus single-round strategies, where each node transmits only one message.

We begin with the problem of computing the Boolean AND function of two variables in Section \ref{sec_and_function}. This problem was studied in \cite{AhlswedeCai}, where it was shown that the exact communication complexity is $\log_{2}3$ bits, for block computation. The lower bound was established using fooling sets, and a novel achievable scheme was presented which minimizes the worst case total number of bits exchanged. The proof technique outlined above can be extended to the problem of computing the AND function of $n$ Boolean variables.   

In Section \ref{sec_threshold_function}, we consider \textit{threshold functions}, which evaluate to $1$ if and only if the total number of $1$s are above a certain threshold. For this class of functions, we devise an achievable strategy which involves each node transmitting in turn using a prefix-free codebook. Further, by intelligent construction of a fooling set, we obtain the exact complexity of computing threshold functions. It is interesting to note that the optimal strategy requires no back-and-forth interaction between nodes. In Section \ref{sec_delta_function}, we also obtain the exact complexity of computing \textit{delta functions} which evaluate to $1$ if and only if there are a certain number of $1$s.

In Section \ref{sec_interval_function}, we study the complexity of computing \textit{interval functions}, which evaluate to 1 if and only if the total number of $1$s belong to a given interval $[a,b]$. For a fixed interval $[a,b]$, the proposed strategy for achievability is order-optimal with optimal preconstant. Additionally, for the interesting class of percentile functions, the proposed single-round strategy is order optimal. The results can be easily extended to the case of many intervals and further, to the case of non-Boolean alphabets.
\section{RELATED WORK}
The problem of worst-case block function computation was formulated in \cite{GiridharKumar}. The authors identify two classes of symmetric functions namely \textit{type-sensitive} functions exemplified by Mean and Median, and \textit{type-threshold} functions, exemplified by Maximum and Minimum. The maximum rates for computation of type-sensitive and type-threshold functions in random planar networks are shown to be $\Theta(\frac{1}{\log n})$ and $\Theta(\frac{1}{\log \log n})$ respectively, for a network of $n$ nodes. A communication complexity approach was used to establish upper bounds on the rate of computation in collocated networks. 

In communication complexity \cite{KushiNisan}, one seeks to minimize the number of bits that must be exchanged between two nodes to achieve worst-case zero-error computation of a function of the node variables. The communication complexity of Boolean functions has been studied in \cite{Wegener}, \cite{OrlitskyElgamal}. Further, one can consider the \textit{direct-sum problem} \cite{KarchmerRazWigderson} where several instances of the problem are considered together to obtain savings. This block computation approach is used to compute the exact complexity of the Boolean AND function in \cite{AhlswedeCai}. In this paper, we considerably generalize this result, which allows us to derive optimal strategies for computing more general classes of symmetric Boolean functions in collocated networks.

While we have considered worst case computation in this paper, one could also impose a probability distribution on the measurements. In \cite{KowshikKumar}, the average complexity of computing a type-threshold function was shown to be $\Theta(1)$, in contrast with the worst case complexity of $\Theta(\log n)$. Thus, we can obtain constant rate computation on the average. 


As argued in \cite{GiridharKumar}, an information-theoretic formulation of this problem combines the complexity of source coding with rate distortion as well as the manifold collaborative possibilities in wireless, together with the complications introduced by the function structure. There is little or no work that addresses this most general framework. One special case, a source coding problem for function computation with side information, has been studied in \cite{OrlitskyRoche}. Recently, the rate region for multi-round interactive function computation has been characterized for two nodes \cite{MaIshwar}, and for collocated networks \cite{MaGuptaIshwar}. 

\section{Zero-error block computation of the AND function}\label{sec_and_function}
\subsection{General problem setting}
Consider a collocated network with nodes $1$ through $n$, where each node $i$ has a Boolean measurement $X_i \in \{0,1\}$. \textit{Every} node wants to compute the same function $f(X_1, X_2, \ldots, X_n)$ of the measurements. We seek to find communication schemes which achieve correct function computation at each node, with minimum worst-case total number of bits exchanged. We allow for the efficiencies of block computation, where each node $i$ has a block of $N$ independent measurements, denoted by $X_i^N$. Throughout this paper, we consider the broadcast scenario, where each node's transmission can be heard by every other node. We also suppose that collisions do not convey information thus restricting ourselves to \textit{collision-free strategies} as in \cite{GiridharKumar}. This means that for the $k^{th}$ bit $b_k$, the identity of the transmitting node $T_k$ depends only on previously broadcast bits $b_1, b_2, \ldots, b_{k-1}$, while the value of the bit it sends can depend arbitrarily on all previous broadcast bits as well as its block of measurements $X_{T_k}^{N}$.

It is important to note that all \textit{interactive} strategies are subsumed within the class of collision-free strategies. A collision-free strategy is said to achieve correct block computation if each node $i$ can  correctly determine the value of the function block $f^N(X_1, X_2, \ldots, X_n)$ using the sequence of bits $b_1, b_2, \ldots$ and its own measurement block $X_i^N$. Let $\mathcal{S}_N$ be the class of collision-free strategies for block length $N$ which achieve zero-error block computation, and let $C(f, S_N, N)$ be the worst-case total number of bits exchanged under strategy $S_N \in \mathcal{S}_N$. The worst-case per-instance complexity of computing a function $f(X_1, X_2, \ldots, X_n)$ is defined by
\begin{displaymath}
C(f) = \lim_{N \rightarrow \infty}\min_{S_N \in \mathcal{S}_N} \frac{C(f, S_N, N)}{N}.
\end{displaymath} 
We call this the \textit{broadcast computation complexity} of the function $f$.

\subsection{Complexity of computing $X_1 \wedge X_2$}
Before we can address the general problem of computing symmetric Boolean functions, we consider the specific problem of computing the AND function, which is $1$ if all its arguments are $1$, and $0$ otherwise. We start by considering just two nodes, namely $1$ and $2$, with measurement blocks $X_1^N$ and $X_2^N$ and we seek to compute the element-wise AND of the two blocks, denoted by $\wedge^N(X_1, X_2)$. This problem was studied in \cite{AhlswedeCai} and we briefly review the proof.
\begin{theorem}\label{thm_two_node_and}
Given any strategy $S_N$ for block computation of $X_1 \wedge X_2$, 
\begin{displaymath}
C(X_1 \wedge X_2, S_N, N) \geq N\log_{2}3.
\end{displaymath}
Further, there exists a strategy $S_N^*$ which satisfies
\begin{displaymath}
C(X_1 \wedge X_2, S_N^*, N) \leq \lceil N \log_{2}3 \rceil .
\end{displaymath}
Thus, the complexity of computing $X_1 \wedge X_2$ is given by $C(X_1 \wedge X_2)=log_{2}3$.
\end{theorem}
\textbf{Proof of achievability:} Suppose node $1$ transmits first using a prefix-free codebook. Let the length of the codeword transmitted be $l(X_1^N)$. At the end of this transmission, both nodes know the value of the function at the instances where $X_1 = 0$. Thus node $2$ only needs to indicate its bits for the instances of the block where $X_1 = 1$. Thus the total number of bits exchanged under this scheme is $l(X_1^N) + w(X_1^N)$, where $w(X_1^N)$ is the number of $1$s in $X_1^N$. For a given scheme, let us define
\begin{displaymath}
L := \max_{X_1^N}(l(X_1^N) + w(X_1^N)),
\end{displaymath} 
to be the worst case total number of bits exchanged. We are interested in finding the codebook which will result in the minimum worst-case number of bits. 

Any prefix-free code must satisfy the Kraft inequality given by $\displaystyle\sum_{X_1^N}2^{-l(X_1^N)} \leq 1$. Consider a codebook with $l(X_1^N) = \lceil N\log_{2}3\rceil - w(x_1^N)$. This satisfies the Kraft inequality since $\sum_{X_1^N}w(X_1^N) = 3^N$. Hence there exists a valid prefix free code for which the worst case number of bits exchanged is $\lceil N\log_{2}3\rceil$, which establishes that $C(X_1 \wedge X_2) \leq \log_{2}3$.

The lower bound is shown by constructing a \textit{fooling set} \cite{KushiNisan} of the appropriate size. We digress briefly to introduce the concept of fooling sets in the context of two-party communication complexity \cite{KushiNisan}. Consider two nodes $X$ and $Y$, each of which take values in finite sets $\mathcal{X}$ and $\mathcal{Y}$, and both nodes want to compute some function $f(X,Y)$ with zero error. 
\begin{definition}[Fooling Set]
A set $E \subseteq \mathcal{X} \times \mathcal{Y}$ is said to be a fooling set, if for any two distinct elements $(x_1, y_1), (x_2, y_2)$ in $E$, we have either 
\begin{itemize}
\item $f(x_1, y_1) \neq f(x_2, y_2)$, or
\item $f(x_1, y_1) = f(x_2, y_2)$, but either $f(x_1,y_2) \neq f(x_1, y_1)$ or $f(x_2,y_1) \neq f(x_1,y_1)$.
\end{itemize}
\end{definition}
Given a fooling set $E$ for a function $f(X_1, X_2)$, we have $C(f(X_1, X_2)) \geq \log_{2}|E|$. We have described two dimensional fooling sets above. The extension to multi-dimensional fooling sets is straightforward and gives a lower bound on the communication complexity of the function $f(X_1, X_2, \ldots, X_n)$. \\
\textbf{Lower bound for Theorem \ref{thm_two_node_and}:} We define the measurement matrix $M$ to be the matrix obtained by stacking the row $X_1^N$  over the row $X_2^N$. Thus we need to find a subset of the set of all measurement matrices which forms a fooling set. Let $E$ the set of all measurement matrices which are made up of only the column vectors {\small{$\{\left[\begin{array}{c} 1 \\ 0 \end{array}\right], \left[\begin{array}{c} 0 \\ 1 \end{array}\right], \left[\begin{array}{c} 1 \\ 1 \end{array}\right]\}$}}. We claim that $E$ is the appropriate fooling set. Consider two distinct measurement matrices $M_1, M_2 \in E$. Let $f^N(M_1)$ and $f^N(M_2)$ be the block function values obtained from these two matrices. If $f^N(M_1) \neq f^N(M_2)$, we are done. Let us suppose $f^N(M_1) = f^N(M_2)$ and since $M_1 \neq M_2$, there must exist one column where $M_1$ has $\left[\begin{array}{c} 0 \\ 1 \end{array}\right]$ but $M_2$ has $\left[\begin{array}{c} 1 \\ 0 \end{array}\right]$. Now if we replace the first row of $M_1$ with the first row of $M_2$, the resulting measurement matrix, say $M^{*}$ is such that $f(M^{*}) \neq f(M_1)$. Thus, the set $E$ is a valid fooling set. It is easy to verify that the $E$ has cardinality $3^N$. Thus, for \textit{any} strategy $S_N \in \mathcal{S}_N$, we must have $C(X_1 \wedge X_2, S_N, N) \geq N\log_{2}3$, implying that $C(X_1 \wedge X_2) \geq \log_{2}3$. This concludes the proof of Theorem \ref{thm_two_node_and}. $\Box$ 
\begin{corollary}\label{cor_two_node_or}
The complexity of the OR function is given by $C(X_1 \vee X_2) = \log_{2}(3)$, since we can view it as $\overline{\overline{X}_1 \wedge \overline{X}_2}$, by deMorgan's laws.
\end{corollary}

The above approach can be easily extended to the general AND function of $n$ variables, and we obtain $C(\wedge(X_1, X_2, \ldots, X_n)) = \log_{2}(n+1)$. We now proceed to provide an exact result for a more general class of functions, called \textit{threshold functions}, which includes AND as a special case.\\
\textbf{Note:} Throughout the rest of the paper, for ease of exposition, we will ignore the fact that terms like $N\log_{2}(n+1)$ may not be integer. Since our achievability strategy involves each node transmitting exactly once, this will result in a maximum of one extra bit per node, and since we are amortizing this over a long block length $N$, it will not affect any of the results.
\section{Complexity of computing Boolean threshold functions}\label{sec_threshold_function}
\begin{definition}[Boolean threshold functions]
A Boolean threshold function $\Pi_{\theta}(X_1, X_2, \ldots, X_n)$ is defined as
\begin{displaymath}
\Pi_{\theta}(X_1, X_2, \ldots, X_n) = \left\{ \begin{array}{l} 1 \quad \textrm{if } \sum_{i}X_i \geq \theta \\ 0 \quad \textrm{otherwise.}\end{array}\right.
\end{displaymath}
\end{definition}
\begin{theorem}\label{thm_bool_threshold}
The complexity of computing a Boolean threshold function is $C(\Pi_{\theta}(X_1, X_2, \ldots X_n)) = \log_{2}\left(\begin{array}{c}n+1 \\ \theta \end{array}\right)$.
\end{theorem}
\textbf{Proof of Achievability:} The upper bound is established by induction on $n$. From Theorem \ref{thm_two_node_and} and Corollary \ref{cor_two_node_or}, the result is true for $n=2$ and for \textit{all} $1 \leq \theta \leq n$, which is the basis step. Suppose the upper bound is true for a collocated network of $(n-1)$ nodes, for all $1 \leq \theta \leq (n-1)$. Given a function $\Pi_{\theta}(X_1, X_2, \ldots, X_n)$ of $n$ variables, consider an achievable strategy in which node $n$ transmits first, using a prefix free codeword of length $l(X_n^N)$. After this transmission, nodes $1$ through $n-1$ can decode the block $X_n^N$. For the instances where $X_n = 0$, these $(n-1)$ nodes now need to compute $\Pi_{\theta}(X_1, X_2, \ldots, X_{n-1})$. For the instances where $X_n = 1$, the remaining $(n-1)$ nodes need to compute $\Pi_{\theta -1}(X_1, X_2, \ldots, X_{n-1})$. From the induction hypothesis, we have optimal strategies for computing these functions. Let $w^{i}(X_n^N)$ denote the number of instances of $i$ in the block $X_n^N$. Under the above strategy, the worst-case total number of bits exchanged is

\vspace{-0.1in}\begin{small}
\begin{displaymath}
L = \max_{X_n^N} \left(l(X_n^N) + w^{0}(X_n^N) \log_{2}\left(\begin{array}{c} n \\ \theta \end{array}\right) + w^{1}(X_n^N) \log_{2}\left(\begin{array}{c} n \\ \theta -1 \end{array}\right) \right). 
\end{displaymath}
\end{small}
\normalsize We want to minimise this quantity subject to the Kraft inequality. Consider a prefix-free codebook which satisfies 
 
\footnotesize 
\begin{displaymath}
l(X_n^N) = N\log_{2}\left(\begin{array}{c} n+1 \\ \theta \end{array}\right) - w^{0}(X_n^N) \log_{2}\left(\begin{array}{c} n \\ \theta \end{array}\right) - w^{1}(X_n^N) \log_{2}\left(\begin{array}{c} n \\ \theta -1 \end{array}\right).
\end{displaymath}
\normalsize This assignment of codelengths satisfies the Kraft inequality since
\begin{eqnarray}
\sum_{X_n^N}2^{-l(X_n^N)} & = & \left(\begin{array}{c}n+1 \\ \theta \end{array}\right)^{-N}\sum_{X_n^N}\left(\begin{array}{c}n \\ \theta \end{array}\right)^{w^{0}(X_n^N)} \left(\begin{array}{c}n \\ \theta -1 \end{array}\right)^{w^{1}(X_n^N)} \nonumber \\
& = & \left(\begin{array}{c}n+1 \\ \theta \end{array}\right)^{-N}\left[\left(\begin{array}{c}n \\ \theta \end{array}\right) + \left(\begin{array}{c}n \\ \theta -1 \end{array}\right)\right]^{N} = 1. \nonumber
\end{eqnarray}
Hence there exists a prefix-free code which satisfies the specified codelengths, and we have $L =   N \log_{2}\left(\begin{array}{c}n+1 \\ \theta \end{array}\right)$, which proves the induction step.\\
\textbf{Proof of lower bound:} We need to find a subset of the set of all $n \times N$ measurement matrices which is a valid fooling set. Consider the subset $E$ of measurement matrices which consist of only columns which sum to $(\theta-1)$ or $\theta$. Since there are $N$ columns, there are $\left[\left(\begin{array}{c}n \\ \theta \end{array}\right) + \left(\begin{array}{c}n \\ \theta -1 \end{array}\right)\right]^{N}$ such matrices. We claim that the set $E$ is a valid fooling set. Let $M_1$, $M_2$ be two distinct matrices in this subset. If $f^{N}(M_1) \neq f^{N}(M_2)$, then we are done. Suppose not. Then there must exist at least one column at which $M_1$ and $M_2$ disagree, say $M_1^{(j)} \neq M_2^{(j)}$. However, both $M_1^{(j)}$ and $M_2^{(j)}$ have the same number of ones. Thus there must exist some row, say $i^*$, where $M_1^{(j)}$ has a zero, but $M_2^{(j)}$ has a one.\\
\textbf{(i)} Suppose $f(M_1^{(j)}) = f(M_2^{(j)}) = 0$. Then, consider the matrix $M_1^*$ obtained by replacing the $i^*$th row of $M_1$ with the $i^*$th row of $M_2$. The $j^{th}$ column of $M_1^*$ has $\theta$ ones, and hence $f(M_1^{*(j)}) = 1$. Hence we have $f(M_1^*) \neq f(M_1)$.\\
\textbf{(ii)} Suppose $f(M_1^{(j)}) = f(M_2^{(j)}) = 1$. Then, consider the matrix $M_2^*$ obtained by replacing the $i^*$th row of $M_2$ with the $i^*$th row of $M_1$. The $j^{th}$ column of $M_2^*$ has $\theta -1$ ones, and hence $f(M_2^{*(j)}) = 1$. Hence we have $f(M_2^*) \neq f(M_2)$.

Thus, the set $E$ is a valid fooling set. From the fooling set lower bound, for \textit{any} strategy $S_N \in \mathcal{S}_N$, we must have $C(\Pi_{\theta}(X_1, X_2, \ldots, X_n), S_N, N) \geq N\log_{2}\left(\begin{array}{c}n+1 \\ \theta \end{array}\right)$ implying that $C(\Pi_{\theta}(X_1, X_2, \ldots, X_n)) \geq \log_{2}\left(\begin{array}{c}n+1 \\ \theta \end{array}\right)$. $\Box$ 
\subsection{Complexity of Boolean delta functions}\label{sec_delta_function}
\begin{definition}[Boolean delta function]
A Boolean delta function $\Pi_{\{\theta\}}(X_1, X_2, \ldots, X_n)$ is defined as:
\begin{displaymath}
\Pi_{\{\theta\}}(X_1, X_2, \ldots, X_n) = \left\{\begin{array}{l} 1 \quad \textrm{if } \sum_{i}X_i = \theta \\ 0 \quad \textrm{otherwise.}\end{array}\right.
\end{displaymath}
\end{definition}
\begin{theorem}\label{thm_bool_delta}
The complexity of computing $\Pi_{\{\theta\}}(X_1, X_2, \ldots, X_n)$ is given by
\begin{displaymath} 
C(\Pi_{\{\theta\}}(X_1, X_2, \ldots, X_n)) = \log_{2}\left[\left(\begin{array}{c}n+1 \\ \theta \end{array}\right) + \left(\begin{array}{c}n \\ \theta + 1 \end{array}\right) \right]. 
\end{displaymath}
\end{theorem}
\textbf{Sketch of Proof:} The proof of achievability follows from an inductive argument as before. The fooling set $E$ consists of measurement matrices composed of only columns which sum up to $\theta -1$, $\theta$ or $\theta + 1$. Thus the size of the fooling set is \begin{displaymath}
\left[\left(\begin{array}{c}n \\ \theta -1 \end{array}\right) + \left(\begin{array}{c}n \\ \theta \end{array}\right) + \left(\begin{array}{c}n \\ \theta + 1 \end{array}\right) \right]^{N}. \Box
\end{displaymath}
\section{Complexity of computing Boolean interval functions}\label{sec_interval_function}
A Boolean \textit{interval function} $\Pi_{[a,b]}(X_1, \ldots, X_n)$ is defined as:
\begin{displaymath}
\Pi_{[a,b]}(X_1, X_2, \ldots, X_n) = \left\{\begin{array}{l} 1 \quad \textrm{if } a \leq \sum_{i}X_i \leq b \\ 0 \quad \textrm{otherwise.}\end{array} \right.
\end{displaymath}
A naive strategy to compute the function $\Pi_{[a,b]}(X_1, \ldots, X_n)$ is to compute the threshold functions $\Pi_{a}(X_1, \ldots, X_n)$ and $\Pi_{b+1}(X_1, X_2, \ldots, X_n)$. However, this strategy gives us more information than we seek, i.e., if $\sum_{i} X_i \in [a,b]^{C}$, then we also know if $\sum_{i} X_i < a$, which is superfluous information and perhaps costly to obtain. Alternately, we can derive a strategy which explicitly deals with intervals, as against thresholds. This strategy has significantly lower complexity.
\begin{theorem}\label{thm_bool_interval}
The complexity of computing a Boolean interval function $\Pi_{[a,b]}(X_1, X_2, \ldots, X_n)$ with $a + b \leq n$ is bounded as follows:
\begin{multline}
\small{
\log_{2}\left[\left(\begin{array}{c}n+1 \\ b+1 \end{array}\right) + \left(\begin{array}{c}n \\ a-1 \end{array}\right)\right] \leq C(\Pi_{[a,b]}(X_1, X_2, \ldots X_n)) }\\
\small{
\leq \log_{2}\left[ \left(\begin{array}{c}n+1 \\ b+1 \end{array}\right) + (b-a + 1)\left(\begin{array}{c}n \\ a-1 \end{array}\right)\right]. \label{int_bound_caseone} } 
\end{multline}
The complexity of computing a Boolean interval function $\Pi_{[a,b]}(X_1, \ldots, X_n)$ with $a +b \geq n$ is bounded as follows:
\begin{multline}
\small{
\log_{2}\left[\left(\begin{array}{c}n+1 \\ a \end{array}\right) + \left(\begin{array}{c}n \\ b+1 \end{array}\right)\right] \leq C(\Pi_{[a,b]}(X_1, X_2, \ldots X_n)) }\\
\small{ \leq \log_{2}\left[ \left(\begin{array}{c}n+1 \\ a \end{array}\right) + (b-a+1)\left(\begin{array}{c}n\\ b+1 \end{array}\right)\right]. \label{int_bound_casetwo} }
\end{multline}
\end{theorem}
\textbf{Proof of lower bound:} Suppose $a+b \leq n$. Consider the subset $E$ of measurement matrices which consist of only columns which sum to $(a-1)$, $b$ or $(b+1)$. We claim that the set $E$ is a valid fooling set. Let $M_1$, $M_2$ be two distinct matrices in this subset. If $f^{N}(M_1) \neq f^{N}(M_2)$, we are done. Suppose not. Then there must exist at least one column at which $M_1$ and $M_2$ disagree, say $M_1^{(j)} \neq M_2^{(j)}$. \\
\textbf{(i)} Suppose $f(M_1^{(j)}) = f(M_2^{(j)}) = 1$. Then, both $M_1^{(j)}$ and $M_2^{(j)}$ have exactly $b$ $1$s. Thus there exists some row, say $i^*$, where $M_1^{(j)}$ has a $0$, but $M_2^{(j)}$ has a $1$. Consider the matrix $M_1^*$ obtained by replacing the $i^*$th row of $M_1$ with the $i^*$th row of $M_2$. The $j^{th}$ column of $M_1^*$ has $(b+1)$ $1$s, and hence $f(M_1^{*(j)}) = 0$, which means $f(M_1^*) \neq f(M_1)$. \\
\textbf{(ii)} Suppose $f(M_1^{(j)}) = f(M_2^{(j)}) = 0$. If both $M_1^{(j)}$ and $M_2^{(j)}$ have the same number of $1$s, then the same argument as in (i) applies. However, if $M_1^{(j)}$ has $(a-1)$ $1$s and $M_2^{(j)}$ has $(b+1)$ $1$s, then there exists some row $i^*$ where $M_1^{(j)}$ has a $0$, but $M_2^{(j)}$ has a $1$. Then, the matrix $M_2^*$ obtained by replacing the $i^*$th row of $M_2$ with the $i^*$th row of $M_1$ is such that $f(M_2^*) \neq f(M_2)$. \\

Thus, the set $E$ is a valid fooling set and $|E| = \left[\left(\begin{array}{c}n \\ b+1 \end{array}\right) + \left(\begin{array}{c}n \\ a -1 \end{array}\right) + \left(\begin{array}{c}n \\ b \end{array}\right)\right]^{N}$. This gives us the required lower bound in (\ref{int_bound_caseone}).

For the case where $a +b \geq n$, we consider the fooling set $E'$ of matrices which are comprised of only columns which sum to $a-1$, $a$ or $b+1$. This gives us the lower bound in (\ref{int_bound_casetwo}).\\
\textbf{Proof of achievability:} Consider the general strategy for achievability where node $n$ transmits a prefix-free codeword of length $l(X_1^N)$, leaving the remaining $(n-1)$ nodes the task of computing a residual function. This approach yields a recursion for computing the complexity of interval functions.

\begin{small}
\begin{displaymath}
C(\Pi_{[a,b]}(X_1, \ldots, X_n)) \leq \log_{2}[2^{C(\Pi_{[a-1, b-1]}(X_1, \ldots, X_{n-1}))} + 2^{C(\Pi_{[a,b]}(X_1, \ldots ,X_{n-1})}].
\end{displaymath}
\end{small}
The boundary conditions for this recursion are obtained from the result for Boolean threshold functions in Theorem \ref{thm_bool_threshold}. We could simply solve this recursion computationally, but we want to study the behaviour of the complexity as we vary $a$, $b$ and $n$. Define $f_{a,b,n} := 2^{C(\Pi_{[a,b]}(X_1, \ldots, X_n))}$. We have the following recursion for $f(a,b,n)$
\begin{equation}\label{f_recursion}
f(a,b,n) \leq f(a-1,b-1,n-1) + f(a,b,n-1). 
\end{equation}
We proceed by induction on $n$. From Theorems \ref{thm_bool_threshold} and \ref{thm_bool_delta}, the upper bounds in (\ref{int_bound_caseone}) and (\ref{int_bound_casetwo}) are true for $n=2$ and all intervals $[a,b]$. Suppose the upper bound is true for all intervals $[a,b]$ for $(n-1)$ nodes. Consider the following cases.\\
\textbf{(i)} Suppose $a + b \leq n-1$. Substituting the induction hypothesis in (\ref{f_recursion}), we get
\begin{eqnarray}
f(a,b,n) & \leq & \left(\begin{array}{c}n \\b \end{array}\right) + (b-a +1)\left(\begin{array}{c}n-1 \\a-2 \end{array}\right) \nonumber \\
& & + \left(\begin{array}{c}n \\b+1 \end{array}\right) + (b-a+1)\left(\begin{array}{c}n-1 \\a-1 \end{array}\right) \nonumber \\
& =  & \left(\begin{array}{c}n+1 \\b+1 \end{array}\right) + (b-a+1)\left(\begin{array}{c}n \\a-1 \end{array}\right). \nonumber
\end{eqnarray}
\textbf{(ii)} Suppose $a+b \geq n+1$. Proof is similar to case (i).\\
\textbf{(iii)} Suppose $a + b = n$. Substituting the induction hypothesis in (\ref{f_recursion}), we get
\begin{eqnarray}
f(a,b,n) & \leq & \left(\begin{array}{c}n \\b \end{array}\right) + (b-a+1)\left(\begin{array}{c}n-1 \\a-2 \end{array}\right) \nonumber \\ 
& & + \left(\begin{array}{c}n \\a \end{array}\right) + (b-a+1)\left(\begin{array}{c}n-1 \\b+1 \end{array}\right) \nonumber \\
& \leq & \left(\begin{array}{c}n+1 \\a \end{array}\right) + (b-a+1)\left(\begin{array}{c}n \\ b+1 \end{array}\right). \nonumber 
\end{eqnarray}
where some steps have been omitted in the proof of the last inequality. This establishes the induction step and completes the proof. $\Box$
\subsection{Discussion of Theorem \ref{thm_bool_interval}}
\textbf{(a)} The gap between the lower and upper bounds in (\ref{int_bound_caseone}) and (\ref{int_bound_casetwo}) is \textit{additive}, and is upper bounded by $\log_{2}(b-a + 2)$ which is $\log_{2}(n+2)$ in the worst case.\\
\textbf{(b)} For fixed $a$ and $b$, as the number of nodes increases, we have $a+b \leq n$ for large enough $n$.  Consider the residual term, $(b-a+1)\left(\begin{array}{c}n \\a-1 \end{array}\right)$ on the RHS in (\ref{int_bound_caseone}). We have
\begin{displaymath}
(b-a+1)\left(\begin{array}{c}n \\a-1 \end{array}\right) = o\left(\left(\begin{array}{c}n+1 \\ b+1 \end{array}\right)\right).
\end{displaymath}
Hence, $C(\Pi_{[a,b]}(X_1, \ldots, X_n)) = \log_{2}\left(\left(\begin{array}{c}n+1 \\ b+1 \end{array}\right)(1 + o(1))\right)$. Thus, for any fixed interval $[a,b]$, we have derived an order optimal strategy with optimal preconstant. The orderwise complexity of this strategy is the same as that of the threshold function $\Pi_{b+1}(X_1, \ldots, X_n)$.  Similarly, we can derive order optimal strategies for computing $C(\Pi_{[n-a,n-b]}(X_1, \ldots, X_n))$ and $C(\Pi_{[a,n-b]}(X_1, \ldots, X_n))$, for fixed $a$ and $b$. \\
\textbf{(c)} Consider a \textit{percentile} type function where $[a,b] = [\alpha n, \beta n]$, with $(\alpha + \beta) \leq 1$. Using Stirling's approximation, we can still show that
\begin{displaymath}
(\beta - \alpha)n\left(\begin{array}{c}n \\ \alpha n -1 \end{array}\right) = o\left(\left(\begin{array}{c}n+1 \\ \beta n+1 \end{array}\right)\right).
\end{displaymath}
Thus we have derived an order optimal strategy with optimal preconstant for percentile functions.\\
\textbf{(d)} Consider the function $f := \Pi_{\cup_{i} [a_i, b_i]}(X_1, \ldots, X_n)$ where the intervals $[a_i, b_i]$ are disjoint, and may be fixed or percentile type. We can piece together the result for single intervals and show that 
\begin{displaymath}
C(f(X_1, \ldots, X_n)) = \log_{2}\left(\sum_{i=1}^{m}g(a_i,b_i,n)(1 + o(1))\right).
\end{displaymath}
\begin{displaymath}
\textrm{where } g(a_i,b_i,n) = \left\{\begin{array}{ll} \left(\begin{array}{c}n+1 \\ b_i+1 \end{array}\right) \textrm{ if } a_i + b_i \leq n\\
\left(\begin{array}{c}n+1 \\ a_i \end{array}\right) \textrm{ if } a_i + b_i \geq n. \end{array} \right.
\end{displaymath}
\section{Concluding remarks}
We have addressed the problem of computing symmetric Boolean functions in a collocated network. We have derived optimal strategies for computing threshold functions and order optimal strategies with optimal preconstant for interval functions. Thus, we have sharply characterized the complexity of various classes of symmetric Boolean functions. Further, since the thresholds and intervals are allowed to depend on $n$, we have provided a unified treatment of type-sensitive and type-threshold functions.

The results can be extended in two directions. First, we can consider non-Boolean alphabets and functions which depend only on $\sum_{i} X_i$. Alternately, we can consider non-Boolean functions of a Boolean alphabet. The fooling set lower bound and the strategy for achievability can be generalized to both these cases. 
\bibliographystyle{unsrt}
\bibliography{itw_biblio}

\begin{thebibliography}{10}

\bibitem{GiridharKumar}
A.~Giridhar and P.~R. Kumar.
\newblock Computing and communicating functions over sensor networks.
\newblock {\em IEEE Journal on Selected Areas in Communication},
  23(4):755--764, April 2005.

\bibitem{AhlswedeCai}
R.~Ahlswede and Ning Cai.
\newblock On communication complexity of vector-valued functions.
\newblock {\em IEEE Transactions on Information Theory}, 40:2062--2067, 1994.

\bibitem{KushiNisan}
E.~Kushilevitz and N.~Nisan.
\newblock {\em Communication Complexity}.
\newblock Cambridge University Press, 1997.

\bibitem{Wegener}
I.~Wegener.
\newblock {\em The Complexity of Boolean Functions}.
\newblock B. G. Teubner, and John Wiley \& Sons, 1987.

\bibitem{OrlitskyElgamal}
A.~Orlitsky and A.~El Gamal.
\newblock Average and randomized communication complexity.
\newblock {\em IEEE Transactions on Information Theory}, 36:3--16, 1990.

\bibitem{KarchmerRazWigderson}
M.~Karchmer, R.~Raz, and A.~Wigderson.
\newblock Super-logarithmic depth lower bounds via direct sum in communication
  coplexity.
\newblock In {\em Structure in Complexity Theory Conference}, pages 299--304,
  1991.

\bibitem{KowshikKumar}
H.~Kowshik and P.~R. Kumar.
\newblock Zero-error function computation in sensor networks.
\newblock In {\em To appear in Proceedings of the 48th IEEE Conference on
  Decision and Control (CDC)}, December 2009.

\bibitem{OrlitskyRoche}
A.~Orlitsky and J.~R. Roche.
\newblock Coding for computing.
\newblock {\em IEEE Transactions on Information Theory}, 47:903--917, 2001.

\bibitem{MaIshwar}
N.~Ma and P.~Ishwar.
\newblock Distributed source coding for interactive function computation.
\newblock {\em Submitted to IEEE Transactions on Information Theory}, 2008.

\bibitem{MaGuptaIshwar}
N.~Ma, P.~Ishwar, and P.~Gupta.
\newblock Information-theoretic bounds for multiround function computation in
  collocated networks.
\newblock {\em Proceedings of the IEEE International Symposium on Information
  Theory (ISIT)}, 2009.

\end{thebibliography}
\end{document}